\documentclass[12pt]{iopart}
\begin{document}

\title{A new type of solution of the Schr\"odinger equation on a self-similar
fractal potential}

\author{N L Chuprikov and O V Spiridonova}

\address{Tomsk State Pedagogical University, 634041, Tomsk, Russia}

\begin{abstract}

Scattering a quantum particle by a self-similar fractal potential on a Cantor set is
investigated. We present a new type of solution of the functional equation for the
transfer matrix of this potential, which was derived earlier from the Schr\"odinger
equation.

\end{abstract}
\pacs{03.65.Ca, 03.65.Xp }


\maketitle

\newcommand{\ppp}{\mbox{\hspace{5mm}}}
\newcommand{\ooo}{\mbox{\hspace{3mm}}}
\newcommand{\ooa}{\mbox{\hspace{1mm}}}

In this Letter we address the model \cite{Ch1,Ch2} of scattering a quantum particle
by a self-similar fractal potential (SSFP) given on a Cantor set. This scattering
problem is, perhaps, the most simple one to allow studying the influence of the
scale invariance of ideal deterministic fractals on physical processes in continuous
media to involve such fractal structures.

Note that the sharp attenuations, found in \cite{Ch1}, in the spectrum of
probability waves transmitted through this ideal fractal potential have also been
observed experimentally (see \cite{Ho1}) in the transmission spectrum of
electromagnetic waves propagating through a real fractal medium (a numerical
modelling for the corresponding pre-fractals see in \cite{Ho2}). However, the
problem is that the model \cite{Ch1,Ch2} remans incomplete in some respects. In this
Letter we present a new type of solution to the Schr\"odinger equation on the SSFP,
in addition to two types presented in \cite{Ch1,Ch2}.

So, let $V(x)$ be a SSFP on the generalized Cantor set in the interval $[0,L]$; each
level of the SSFP consists of $N$ ($N\geq2$) SSFPs of the next level whose width is
$\alpha$ times smaller than that of the former (see \cite{Ch2}). Let also $W$ be a
power of the SSFP, that is, its total area: $W=\int_{-\infty}^\infty V(x)dx$. In
line with \cite{Ch1,Ch2}, for a particle with a given energy $E$
($E=\hbar^2k^2/2m$), the transfer matrix $Z(\phi)$ ($\phi=kL)$ of the SSFP must obey
the functional equation

\begin{equation} \label{1}
\fl {\bf Z}(\phi)={\bf Z}(\alpha\phi) \left[{\bf D}(\gamma\phi){\bf
Z}(\alpha\phi)\right]^{N-1};
\end{equation}
\begin{equation*}
\fl {\bf Z}(\phi)=\left(\begin{array}{cc} q(\phi) & p(\phi) \\
p^*(\phi) & q^*(\phi) \end{array} \right), \hspace{8mm} {\bf
D}(\phi)=\left(\begin{array}{cc} e^{i\phi} & 0 \\ 0 & e^{-i\phi}
\end{array} \right);
\end{equation*}
\[\fl
q(\phi)=\frac{1}{\sqrt{T(\phi)}}\exp\left[-iJ(\phi)\right], \hspace{8mm}
p(\phi)=\sqrt{\frac{R(\phi)}{T(\phi)}}\exp\left[i\left(\frac{\pi}{2}+F(\phi)\right)\right];
\]
$\gamma=\frac{\alpha-N}{\alpha(N-1)}$; $R=1-T$; $T(\phi)$, $J(\phi)$ and $F(\phi)$
are, respectively, the transmission coefficient and phase characteristics of the
SSFP; here $F=0$ for the SSFP-barriers and $F=\pi$ for the SSFP-wells (see
\cite{Ch2}).

As it has turned out, Eq. (\ref{1}) not uniquely determines the transfer matrix of
the SSFP. Two different types of solutions of this equation have been presented in
\cite{Ch1,Ch2}. Remind that the first type was obtained for any values of $W$,
$\alpha$ and $N$ to characterize the SSFP. In this case, for small values of $\phi$,
$\sqrt{T(\phi)}\sim y(\phi)\sim \phi^s$, where $s$ is the fractal dimension of the
Cantor set: $s=\ln(N)/\ln(\alpha)$; $y=\frac{\pi}{2}-J$. The second type of
solutions exists only for the SSFP-barriers, if $W=\frac{3N\hbar^2}{mL}.$ For this
type, $\sqrt{T(\phi)}\sim y(\phi)\sim \phi$ for small values of $\phi$.

In this Letter we present a new (third) type of solutions (found by Chuprikov), with
a cardinally different behavior of the tunneling parameters in the asymptotic
region. Namely, in this case, for small values of $\phi$ we have
\begin{eqnarray} \label{2}
\fl T(\phi)=\left\{1+\cosh^2[\omega(\ln(\phi))]\sinh^2(c\phi^{-s})\right\}^{-1},\nonumber \\
\fl J(\phi)=\arctan\{\sinh[\omega(\ln(\phi))]\tanh(c\phi^{-s})\},
\end{eqnarray}
where $c$ is a nonzero constant; $\omega$ is a nonzero real-valued function to obey
the condition, $\omega[\ln(\phi)]=\omega[\ln(\phi)+\ln(\alpha)].$

To extend this solution onto the whole $\ln(\phi)$-axis, one has to use the recurrence
relations (18) and (19) presented in \cite{Ch2}. As in \cite{Ch1,Ch2}, we display here
$\ln(R/T)$ versus $\ln(\phi)$. Figures (1)-(6) show this function for three values of
$\omega$ - 1, 10 and 15 - and three values of $c$ - 0.001, 0.01 and 0.1. As is seen,
there are three regions on the $\ln(\phi)$-axis, with a qualitatively different
dependence of $\ln(R/T):$

\noindent in the left region
\[\fl \ln\ln\left(R/T\right) \sim \ln(2|c|)-s \ln(\phi);\]
in the right one \[\fl \ln(\widetilde{R/T}\ddot{}\hat{}) \sim -2s \ln(\phi),\] where
$\widetilde{R/T}$ is the envelop of $R(\phi)/T(\phi)$.

As regards the middle region, two qualitatively different types of changing this
function are possible here. For $\omega=1$ and all three values of $c$ (see figures
(1) and (2))
\[\fl \ln\left(R/T\right) \sim -2s \ln(\phi).\] Such behavior also occurs for
$\omega=10$ and $c=0.001$ (see figures (3) and (4)). At the same time, for $\omega=15$
and all three values of $c$ (see figures (5) and (6)), as well as for $\omega=10$ and
$c=0.1$ (see figures (3) and (4)) we have
\[\fl \ln\left(R/T\right) \sim -2 \ln(\phi).\]

Note that the right region appears for all three types of solutions (see \cite{Ch2}).
As regards the left one to follow from (\ref{2}), such a behavior is a distinctive
feature of the third type of solutions.

It is also important to note here that for the solutions of the first and third types
the phase path of the wave inside the out-of-barrier regions (i.e., in the regions
where the potentials are equal to zero) is infinitesimally small in comparison with
the wave path in the barrier regions. This feature distinguishes these types of
solutions from the second one.

A simple analysis shows that the tunneling parameters are non-differentiable functions
at the point $\phi=0$, when $\omega$ depends on $\phi$. Figure 7 shows the function
$R(\phi)/T(\phi)$ for this case.

So, there are at least three types of the transfer matrices of the SSFP. As is seen,
though all of them are nonzero only on the Cantor set, i.e., the set of zero
measure, we deal with different potentials. The Cantor set is a non-countable one,
and, thus, it yet provides a much enough room for setting potentials with such
different scattering properties.

Of course, in this case, it is of great importance is to find the sequences of
pre-fractals to lead to the SSFPs, when the generation number of pre-fractals tends
to infinity. Additionally, another open question regarding the model is that the
parameters to enter the third type of solutions remain to be connected to the SSFP
parameters.

\section*{Figure captions}
\begin{verbatim}
\Figure{\label{fig1}The $\ln(\phi)$-dependence of $\ln(R/T)$ for $s=0.5$, $c=0.001$
and $\omega=1$; bold full curve - $N=2$, thin full curve - $N=4$; points show the
asymptote $13-2s\ln(\phi)$.
\end{verbatim}

\begin{verbatim}
\Figure{\label{fig2}The $\ln(\phi)$-dependence of $\ln(R/T)$ for $N=3$, $\alpha=13$
and $\omega=1$; broken curve - $c=0.1$; thin full curve - $c=0.01$; bold full curve -
$c=0.001$; circles show the asymptote $10-2s\ln(\phi)$.
\end{verbatim}

\begin{verbatim}
\Figure{\label{fig3}The $\ln(\phi)$-dependence of $\ln(R/T)$ for $s=0.5$, $c=0.001$
and $\omega=10$; bold full curve - $N=2$, thin full curve - $N=4$; points show the
asymptote $3-2s\ln(\phi)$.
\end{verbatim}

\begin{verbatim}
\Figure{\label{fig4}The $\ln(\phi)$-dependence of $\ln(R/T)$ for $N=3$, $\alpha=13$
and $\omega=10$; broken curve - $c=0.1$; thin full curve - $c=0.01$; bold full curve -
$c=0.001$; points show the asymptote $8-2\ln(\phi)$; circles show the asymptote
$4-2s\ln(\phi)$.
\end{verbatim}

\begin{verbatim}
\Figure{\label{fig5}The $\ln(\phi)$-dependence of $\ln(R/T)$ for $s=0.5$, $c=0.001$
and $\omega=15$; bold full curve - $N=2$, thin full curve - $N=4$; points show the
asymptote $5-2\ln(\phi)$.
\end{verbatim}

\begin{verbatim}
\Figure{\label{fig6}The $\ln(\phi)$-dependence of $\ln(R/T)$ for $N=3$, $\alpha=13$
and $\omega=15$; broken curve - $c=0.1$; thin full curve - $c=0.01$; bold full curve -
$c=0.001$; points show the asymptote $8-2\ln(\phi)$; circles show the asymptote
$4-2s\ln(\phi)$.
\end{verbatim}

\begin{verbatim}
\Figure{\label{fig7}The $\ln(\phi)$-dependence of $\ln(R/T)$ for $s=0.5$, $c=0.001$
and $\omega=15\left[\sin\left(2\pi\frac{\ln(\phi)}{\ln(\alpha)}\right)+1.001\right]$;
bold full curve - $N=2$, thin full curve - $N=4$; points show the asymptotes
$5-2\ln(\phi)$ and $14-2s\ln(\phi)$.
\end{verbatim}

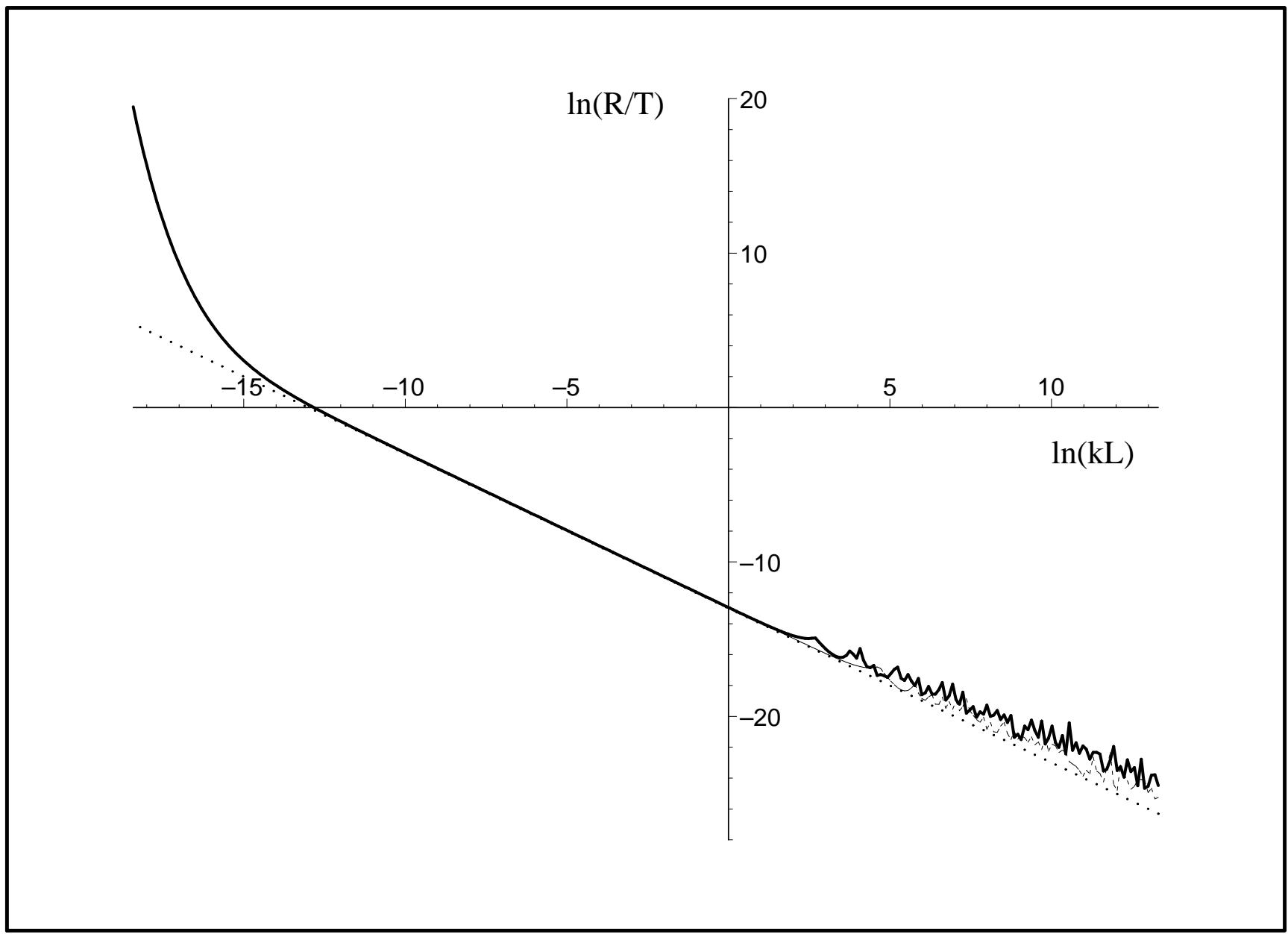
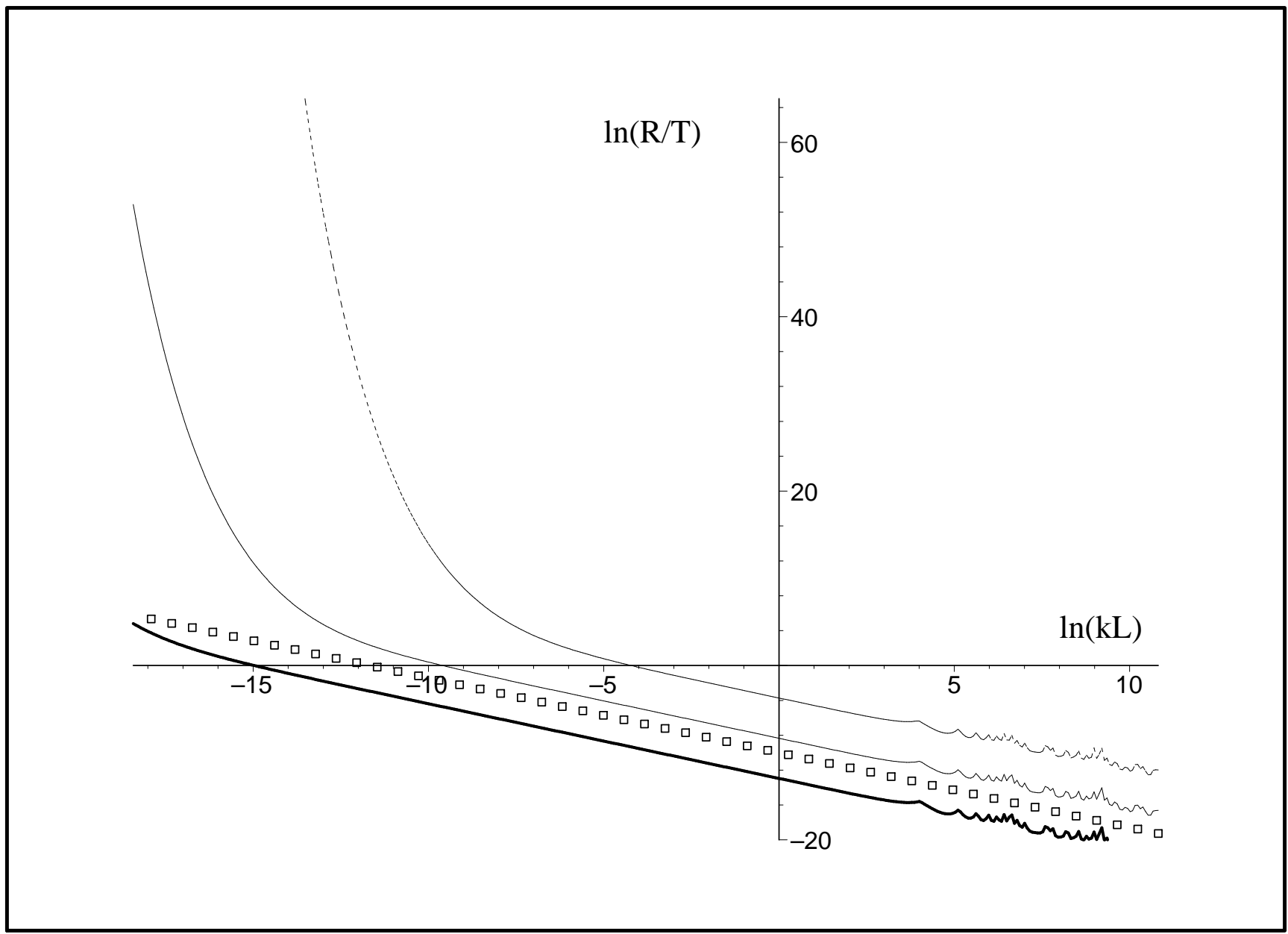
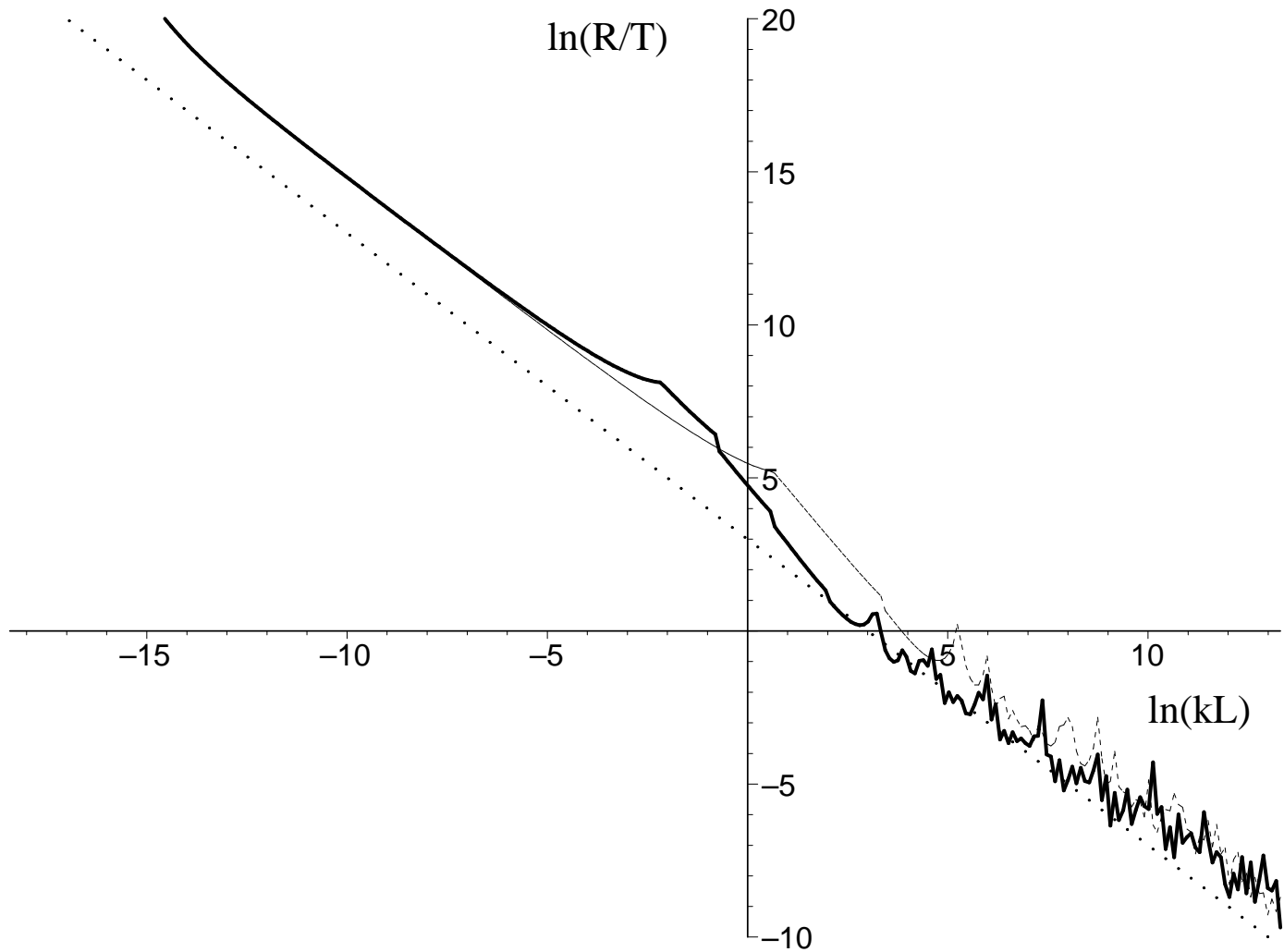
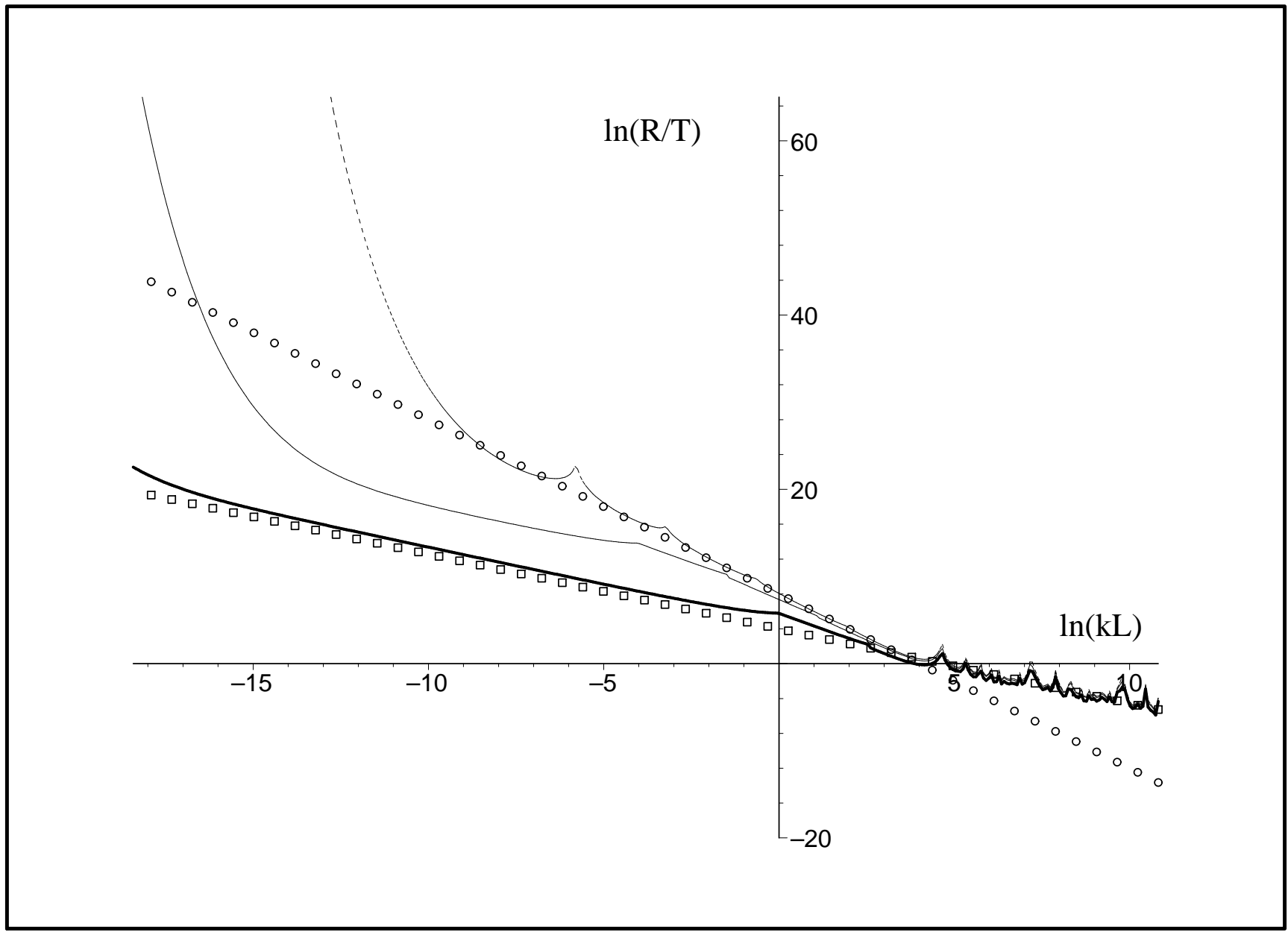
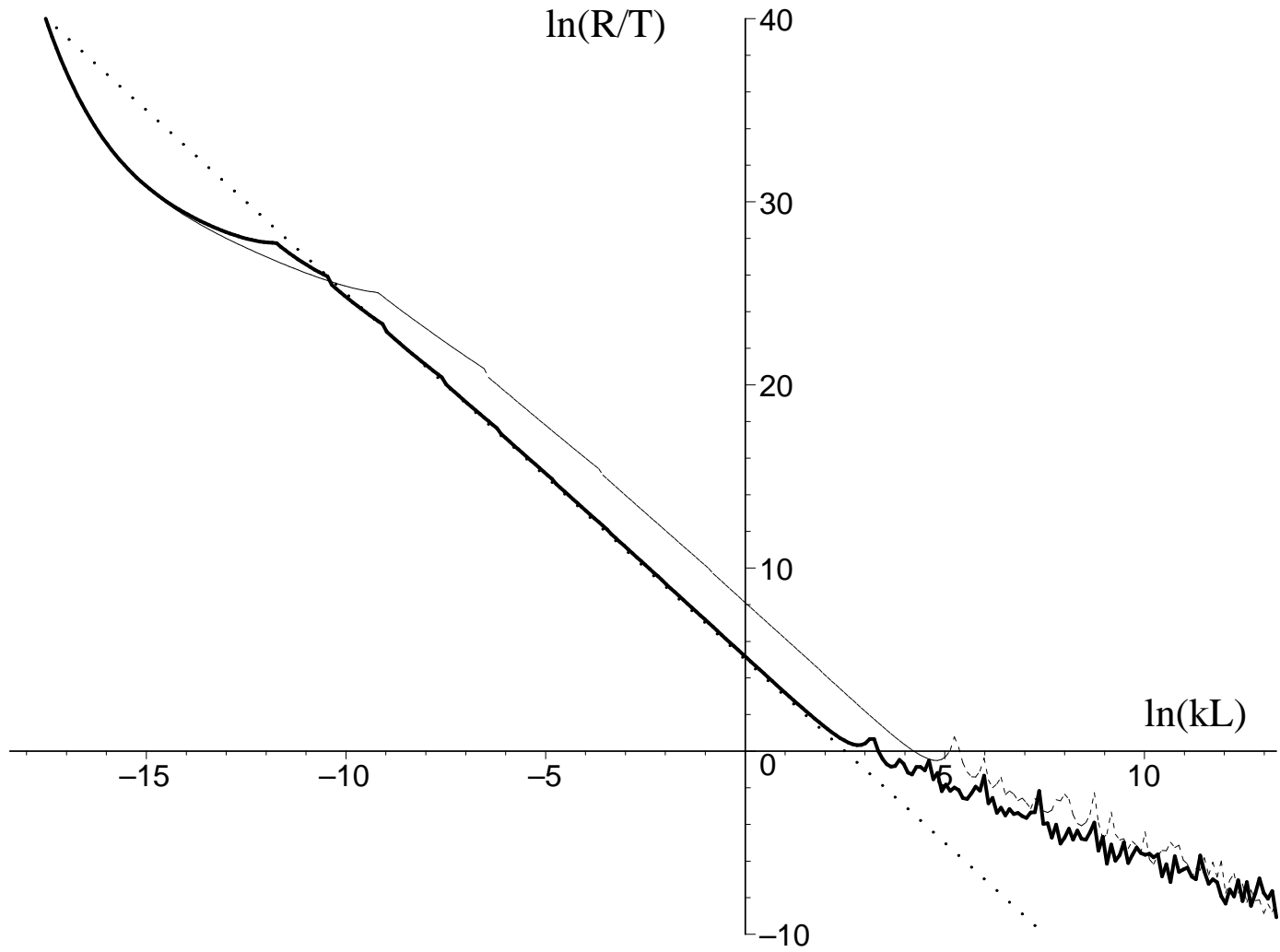
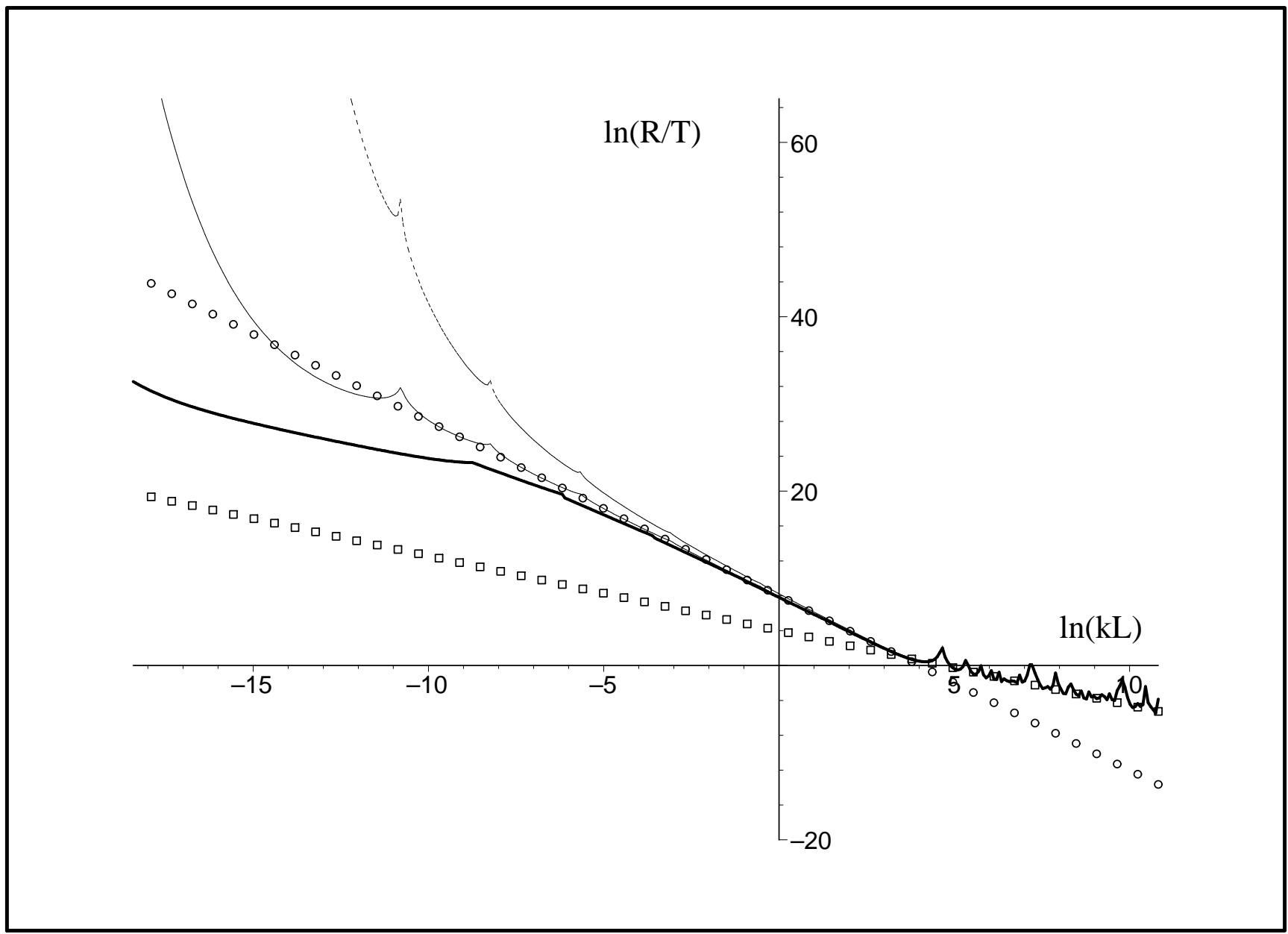
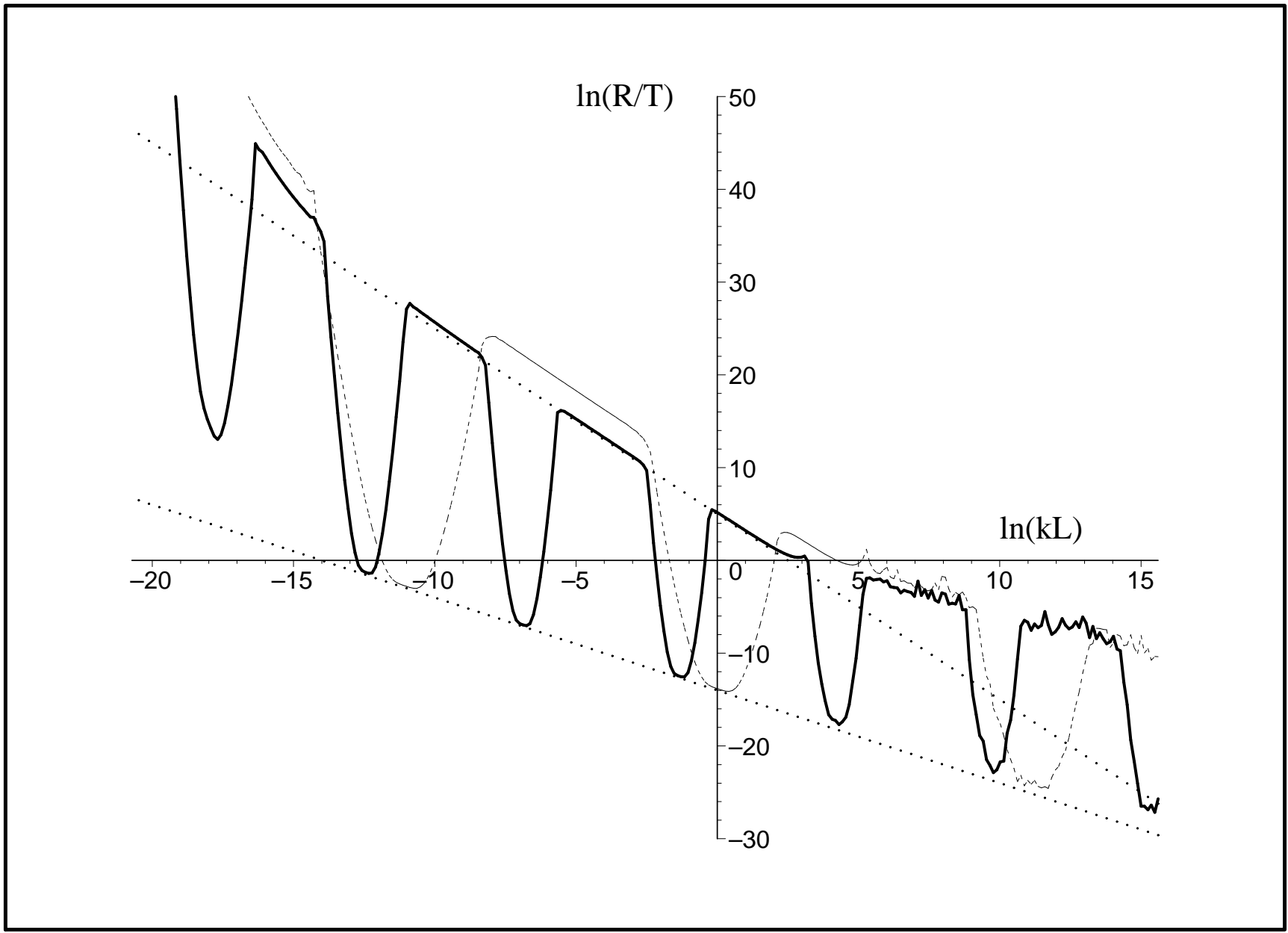
\end{document}